\documentclass[a4paper,10pt,onecolumn,nofootinbib,aps,prd]{revtex4-2}
\usepackage[latin1]{inputenc}
\usepackage{amsmath}
\usepackage{amsfonts}
\usepackage{amssymb,amsthm}
\usepackage{graphicx}
\usepackage[left=2.5cm, right=2.5cm, top=3cm, bottom=3cm]{geometry}
\usepackage{float}
\usepackage{tikz}
\usetikzlibrary{shapes,arrows,shadings,shadows}
\usepackage{graphicx}
\usepackage[compat=1.1.0]{tikz-feynman} 
\usepackage{hyperref}
\usepackage[english]{babel}
\usepackage{subcaption}
\usepackage[justification=centering,singlelinecheck=false]{subcaption}
\usepackage[justification=RaggedRight,singlelinecheck=false]{caption}
\usepackage{amsfonts}
\usepackage{braket}
\usepackage{mathrsfs}
\usepackage{empheq}
\usepackage[shortlabels]{enumitem}
\usepackage{pgfplots}
\usepackage{bigints}
\usepackage{tikz, pgf}
\usetikzlibrary{decorations.pathmorphing}
\usepackage{afterpage}
\usetikzlibrary{arrows.meta}
\tikzset{%
  >={Latex[width=2mm,length=2mm]},
            base/.style = {rectangle, rounded corners, draw=black,
                           minimum width=4cm, minimum height=1cm,
                           text centered, font=\sffamily},
}
\usetikzlibrary{calc,bending,decorations.markings}
\usepackage{setspace}

\usepackage{graphicx}
\usepackage{dcolumn}
\usepackage{bm}

\begin{document}
 	
\title{Leptogenesis with triplet scalars at electroweak scale}
 	
\author{Rituparna Ghosh}
\email{rg20rs072@iiserkol.ac.in}
\affiliation{Department of Physical Sciences \\ Indian Institute of Science Education and Research Kolkata \\ Mohanpur, Nadia - 741246, India}%
\author{Biswarup Mukhopadhyaya}%
\email{biswarup@iiserkol.ac.in}
\affiliation{Department of Physical Sciences \\ Indian Institute of Science Education and Research Kolkata \\ Mohanpur, Nadia - 741246, India}%
\author{Utpal Sarkar}%
\email{utpal.sarkar.prl@gmail.com}
\affiliation{Department of Physical Sciences \\ Indian Institute of Science Education and Research Kolkata \\ Mohanpur, Nadia - 741246, India}%

\begin{abstract}
Up until now the works regarding leptogenesis have discussed different mechanisms to explain the observed baryon asymmetry of the universe (BAU). The type-$II$ seesaw mechanism employing triplet scalars has been well studied in this context. In this work we have invoked an extended Georgi-Machacek type scenario, capable of generating BAU through leptogenesis while keeping the triplet VEV at a moderate range where it plays a nontrivial role in electroweak symmetry breaking (EWSB). The effect of mixing between different $SU(2)$ scalar multiplets have been considered and the decaying scalar mass responsible for the generation of the observed BAU has been kept within a TeV and hence one can probe this possibility at the Large Hadron Collider (LHC) itself.  

\end{abstract}
\maketitle
\newpage
\tableofcontents
\newpage

\section{Introduction}

It is well-known from various observations that the ratio of baryon number density to photon number number density in our universe is about\cite{wmap}-\cite{Pl2}
\begin{equation}
\frac{n_B}{n_\gamma } \approx 6.1 \times 10^{-10} \, .
\end{equation}
This observed BAU poses the question as to how such an asymmetry developed from a B-symmetric beginning?

Sakharov established three conditions as essential for the generation of baryon asymmetry out of a $B = 0$ initial condition\cite{Sakharov} : 
\begin{enumerate}
	\item $B$ - violating interaction.
	\item $C$ and $CP$ violation  
    \item Departure from thermal equilibrium.
	\end{enumerate}

 Early models to generate baryon asymmetry were mostly based on the grand unified theories (GUT) which generate the $B+L$ asymmetry \cite{yanagida}-\cite{so10}.  Since baryon number $B$ and lepton number $L$ are not good symmetries of the SM once the higher order effects are taken into account, the net baryon number observed at BBN scale is a nontrivial combination of $B+L$ and $B-L$ asymmetries. Here one must note that $B+L$ violating sphaleron transition plays a crucial role in generating the observed BAU \cite{Kuzmin}\cite{Rubakov1996}. The sphaleron transition occurs very fast in the temperature range $10^{12}$~ GeV to 100-200 GeV and washes out all the asymmetry generated in $B+L$ by any other mechanism. Therefore, one can justify the observed BAU, if the asymmetry is generated practically via $B-L$ violating interactions. During the $B-L$ conserving sphaleron transition, this asymmetry translates into baryon asymmetry which should be compared to the observed BAU\cite{Flanz-sarkar}-\cite{blus}.  

Before going into further details on baryogenesis, here we briefly discuss a few facts on the thermal history of our universe starting from, say, the Grand Unification scale. As already mentioned, all the $B+L$ asymmetry generated by GUT get washed out by sphaleron. Around the electroweak scale, the sphaleron transition slows down and tends to go out of equilibrium. Now if any particle generates $B+L$ asymmetry at this scale, that may not get completely washed out due to sphaleron. 

This opens up a new way to look at baryogenesis and many studies have been done in this direction, where one first generates a lepton asymmetry through $L$- violating interactions and then a part of the lepton asymmetry gives birth to the observed BAU via sphaleron effects. Additionally, most of the models that offer baryogenesis via leptogenesis predict a neutrino mass naturally thus explaining the neutrino oscillation data. Fukugida and Yanagida first proposed a model in this direction where a lepton asymmetry is been created by the lepton number and CP violating decay of a heavier right handed neutrino (RHN) \cite{Fukugita}. A model of leptogenesis was later developed by Ma and Sarkar where a lepton asymmetry is generated from the decay of a heavier scalar triplet \cite{Mas}. 

In the frameworks where asymmetry is created by the interference between tree level and vertex correction diagram for lepton number violating decays, the required mass scale of the decaying scalar is around $10^{12}$ GeV \cite{Davidson2002}. Consequently, the scenario could not be tested in a terrestrial experiment. But once along with these two diagrams, the self-energy diagrams of decaying scalar, including that mediated by another near-degenerate scalar, is taken into account, the asymmetry could be increased near unity due to a resonant effect. This can happen even when masses of the scalars are as small as a TeV or less. Such a situation admits of collider signatures, thus implying the possibility of leptogenesis being tested in terrestrial experiments \cite{Pilaftsis1997}-\cite{chen}.

In this work we focus on a scenario similar to that proposed by Ma and Sarkar\cite{Mas} where the asymmetry is enhanced by a resonant effect. This scenario consists of two complex scalar triplets both of which couple to leptons via $\Delta L = 2 $ lepton number violating coupling and consequently generates a Mojorana mass for the neutrinos. Before the EWPT, decays of the triplet Higgs scalars generate very large B and L asymmetry through resonant mechanisms, but (B+L) part of the asymmetry is washed out due to sphaleron interactions. The (B-L) combination of the  asymmetry gets converted to the required baryon asymmetry of the universe by the fast (B+L) sphaleron processes. After the EWPT, (B+L) violating sphaleron processes become weaker and weaker and they can affect the (B+L) only partially, and the remaining (B+L) asymmetry provides us with the required baryon asymmetry of the universe. 

However here we will look at this scenario, in a slightly wider perspective. The questions we are asking here are,
\begin{enumerate}
    \item Can leptogenesis scale be lowered by the resonant effect? If so, to what extent?
    \item How much triplet VEV is consistent with leptogenesis, neutrino mass and collider data simultaneously?
\end{enumerate}

 In general, non-doublet VEV faces a severe constraint imposed by the $\rho$-parameter and cannot play much significant role in electroweak symmetry breaking. So, naturally it becomes interesting to ask the question 2 above. Moreover, depending on the values of the triplet VEV, the collider signature could change significantly, so one has to exhaust all possible ranges of this VEV in order to obtain a full picture. Keeping this in mind, we take up a more generalized version of a well-known scenario, namely, the GM model which can lead to resonant leptogenesis. We test the viability of this proposal, with the triplet VEV ranging from small values to tens of GeV, and neutrino masses of the order of $10^{-2} - 10^{-1}$ eV.

 The motivation for using an extended GM scenario is that it admits of situations where the triplets can have VEV of the order of at least tens of GeV. However, after moderate scanning of the parameter space, we find that the regions compatible with the observed BAU as well as with collider data gets rather restricted.

The paper has been organized as follows: first we discuss the method of generating baryon asymmetry from a lepton asymmetry in section \ref{RL}. In the same section we also provide a brief discussion of resonant leptogenesis. The theoretical scenario of interest has been discussed in section \ref{theo}. We reveal our results in section \ref{res} and we finally summarize and conclude in section 5 \ref{cncl}.

\section{A brief review of baryogenesis through leptogenesis}
\label{RL}
The SM at the tree level conserves baryon number (B) and lepton number (L). However, upon inclusion of higher-order effects, both the $B$ and $L$ currents receive non-zero divergences. As a consequence of this anomaly, the baryon number changes as the universe cools down. Baryon number violation is a critical component of any theory that seeks to explain the matter-antimatter asymmetry. But even in the absence of baryon number violating interaction, an asymmetry in $B-L$ can also successfully address the observed BAU as sphaleron transition converts $B-L$ asymmetry to $B$-asymmetry. Hence, sphaleron transition plays a crucial role in explaining observed BAU.

Sphaleron transitions are non-perturbative processes in the SM that violate baryon and lepton number while conserving the difference $B-L$. These transitions occur at high temperatures, such as those present in the early universe, and play a crucial role in baryogenesis scenarios. In particular, they can convert a lepton asymmetry into a baryon asymmetry, making them essential for mechanisms like leptogenesis. Sphaleron processes become efficient above the electroweak phase transition temperature but are suppressed at lower temperatures. For more details, the reader is referred to \cite{Kuzmin}\cite{Rubakov1996}

\subsection{Resonant leptogenesis}

In this work we have considered the CP violating decay of a heavier scalar state to leptons as the source of CP violation. As a result, when the scalar decays in out of equilibrium condition, a net lepton number is generated. This leads to leptogenesis and the asymmetry thus generated is then transferred to the baryon sector via the sphaleron transition. With a hierarchial mass spectrum of the scalar sector, succesful baryogenesis occurs when decaying scalar mass is around $10^{10}$ GeV . In the contrary when two scalar states with lepton number violating and CP violating interactions become quasi-degenerate, the resulting $CP$ asymmetry is resonantly enhanced. This mechanism is known as resonant leptogenesis, that offers a way to enhance the level of CP violation and serves to lower the required mass scale of the decaying particles.

In this work, we consider a model with two heavy scalar states, $\chi_1^{\pm\pm}$ and $\chi_2^{\pm\pm}$, both belonging to complex SU(2) triplets, which are nearly degenerate in mass i.e $|m_1 - m_2| \le \frac{\Gamma_2}{2}$ with $m_1$ and $m_2$ being the masses of $\chi_1^{\pm\pm}$ and $\chi_2^{\pm\pm}$ respectively and $\Gamma_2$ is the decay width of $\chi_2^{\pm\pm}$. The lagrangian is given by,

\begin{eqnarray}
\label{ltype2}
-\mathcal{L} &=& \sum_{a=1,2} \Big[ M_a^2 \chi_a^{\dagger} \chi_a + y_{aij} \Big( \chi_a^0 \nu_i \nu_j + \frac{\chi_a^+}{\sqrt{2}} (\nu_i l_j + \nu_j l_i) + \chi_a^{++} l_i l_j \Big) \nonumber \\
& & \quad  + \mu_a \Big( \chi_a^0 \phi^0 \phi^0 + \chi_a^+ \phi^0 \phi^- + \chi_a^{++} \phi^- \phi^- \Big) + \text{h.c.} \Big]
\end{eqnarray}

where $a$ is the index listing the triplet and $y_{aij}$ is the corresponding $\Delta L = 2$ Yukawa matrix element.

\begin{figure}[htb!]
     \begin{subfigure}[h]{0.48\textwidth}
         \centering
         \includegraphics[width=0.8\textwidth,height=0.8\textwidth]{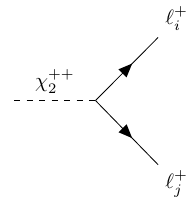}
         \label{feyn1}
     \end{subfigure}
     \hfill
     \begin{subfigure}[h]{0.48\textwidth}
         \centering
    \includegraphics[width=1.1\textwidth,height=0.8\textwidth]{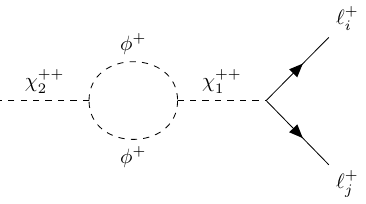}
         \label{feyn2}
    \end{subfigure}
     \caption{Tree level and one loop level Feynman diagrams that participate in resonant leptogenesis}
     \label{feyn}
\end{figure}

The diagrams that contribute to generate $CP$ asymmetry is shown in Fig. \ref{feyn}. In addition to these two diagrams, vertex correction diagram also takes part in generating lepton asymmetry but that has a sub-dominant effect with respect to these two. In resonant leptogenesis interference between the two amplitudes of Fig. \ref{feyn} enhances the $CP$ asymmetry.

The thus generated $CP$ asymmetry mathematically quantified in terms of $\epsilon$:

\begin{equation}
\label{ep}
    \epsilon = 2\frac{\Gamma(\chi_2^{--} \to \ell^{-}\ell^{-}) - \Gamma(\chi_2^{++} \to \ell^{+}\ell^{+})}{\Gamma(\chi_2^{--}) + \Gamma(\chi_2^{++})}
\end{equation}

Follwing Ref. \cite{Mas}, in terms of potential parameters $\epsilon$ is given by,
\begin{equation}
\label{ep}
    \epsilon = \frac{Im[\mu_1\mu_2^*\Sigma_{k,l}y_{1kl}y_{2kl}^*]}{16\pi^2(m_1^2-m_2^2)}\frac{m_2}{\Gamma_2}
\end{equation}

Note that, we have considered $\chi_1^{\pm\pm}$ to be slightly heavier than $\chi_2^{\pm\pm}$.

As emphasized earlier one of the primary advantages of resonant leptogenesis is that it reduces the need for an extremely high scale for the decaying particles as required for the leptogenesis scenarios in hierarchical particle spectrum. This opens up the possibility that the scale of leptogenesis scale could be sufficiently low , so as to be potentially detectable in accelerator experiments. Moreover the enhanced CP violation in resonant leptogenesis, driven by the nearly degenerate triplet scalars $\chi_1^{\pm\pm}$ and $\chi_2^{\pm\pm}$, leads to a more efficient transfer of the lepton asymmetry to  baryon asymmetry via sphaleron transition, even in situations where traditional leptogenesis might fail.

\section{The theoretical framework}
\label{theo}

The lagrangian in eq. \eqref{ltype2} provides the minimalistic framework that explains the observed BAU through leptogenesis. But as already emphasized, one of the major objectives of this study is to quantify the maximum triplet VEV that satisfies the constraints imposed by BAU, neutrino mass, and collider data. Hence, extending the Type-$II$ seesaw scenario by an additional complex triplet will not serve the purpose as the $\rho$ parameter will kill the triplet contribution to the EWSB. For this reason, we are taking up an extended GM-type scenario, where the scalar sector of the GM model has been extended by another complex triplet scalar. This ensures that the triplet VEV is relatively less constrained from the constraint imposed by the $\rho$- parameter.

The scalar sector of the model consists of a doublet $\phi$ , a real triplet $\xi$, and two complex scalar triplet $\chi_1$ and $\chi_2$, given by,

\begin{equation}
	\phi=\begin{pmatrix}
		\phi^+ \\
		\phi^0
	\end{pmatrix},~
	\chi_1 = \begin{pmatrix}
		\frac{\chi_1^+}{\sqrt{2}} & -\chi_1^{++} \\
		\chi_1^0 & -\frac{\chi_1^+}{\sqrt{2}}
	\end{pmatrix} , ~
    \chi_2 = \begin{pmatrix}
		\frac{\chi_2^+}{\sqrt{2}} & -\chi_2^{++} \\
		\chi_2^0 & -\frac{\chi_2^+}{\sqrt{2}}
	\end{pmatrix} ~ , ~\xi = \begin{pmatrix}
		\frac{\xi^0}{\sqrt{2}} & -\xi^+ \\
		-\xi^- & -\frac{\xi^0}{\sqrt{2}}
	\end{pmatrix}, 
\end{equation}
	
In terms of $\phi,\chi_1,\chi_2$ and $\xi$, the scalar potential of the scenario is given by\cite{chen}
\begin{eqnarray}
\label{pot}
V(\phi,\chi_1,\chi_2,\xi) &=& m_\phi^2 (\phi^\dagger \phi) + m_\xi^2\text{Tr}(\xi^2) +\mu_{\phi\xi}\phi^\dagger \xi\phi + \lambda (\phi^\dagger \phi)^2 +\rho_3[\text{Tr}(\xi^2)]^2
+ \sigma_3\text{Tr}(\xi^2)\phi^\dagger \phi 
\notag\\
&+& \Sigma_{i=1,2}  [ m_{\chi_{i}}^2 \text{Tr}(\chi_{i}^\dagger\chi_{i}) +\rho_{1 \chi_i} [\text{Tr}(\chi_{i}^\dagger\chi_{i})]^2+\rho_{2 \chi_i} \text{Tr}(\chi_{i}^\dagger \chi_{i}\chi_{i}^\dagger \chi_{i})
\notag\\
&+& \rho_{4 \chi_i} \text{Tr}(\chi_{i}^\dagger\chi_{i})\text{Tr}(\xi^2) + \rho_{5 \chi_i} \text{Tr}(\chi_{i}^\dagger \xi)\text{Tr}(\xi \chi_{i}) + \sigma_{1 \chi_i} \text{Tr}(\chi_{i}^\dagger \chi_{i})\phi^\dagger \phi+\sigma_{2 \chi_i} \phi^\dagger \chi_{i}\chi_{i}^\dagger \phi 
 \notag\\
&+& \left( \mu_{i}^{\prime} \phi^\dagger \chi_{i} \tilde{\phi} + \text{H.c.} \right) + \left( \sigma_{4 \chi_i} \phi^\dagger \chi_{i}\xi \tilde{\phi} + \text{H.c.} \right) ] \notag\\
&+& \Sigma_{i,j=1,2 , i\ne j}  [ m_{\chi_{ij}}^2 \text{Tr}(\chi_{i}^\dagger\chi_{j})  +\rho_{1 \chi_{ij}} [\text{Tr}(\chi_{i}^\dagger\chi_{j})]^2+\rho_{2 \chi_{ij}} \text{Tr}(\chi_{i}^\dagger \chi_{j}\chi_{i}^\dagger \chi_{j})
\notag\\
&+& \rho_{4 \chi_i} \text{Tr}(\chi_{i}^\dagger\chi_{j})\text{Tr}(\xi^2) + \rho_{5 \chi_i} \text{Tr}(\chi_{i}^\dagger \xi)\text{Tr}(\xi \chi_{j}) + \sigma_{1 \chi_i} \text{Tr}(\chi_{i}^\dagger \chi_{j})\phi^\dagger \phi \notag \\
&+& \sigma_{2 \chi_i} \phi^\dagger \chi_{i}\chi_{j}^\dagger \phi + \text{H.c} ]\label{pot_gen}
\end{eqnarray}

However, we have used a simplified form of the scalar potential in our illustrative study. The assumptions we used are outlined at the beginning of the next section.

The scalar sector of the theory includes two doubly charged scalars, four singly charged scalars (one of which is the charged Goldstone mode), and seven neutral scalars, one of which is absorbed as the longitudinal mode of the $Z$-boson.

We assume that the mixing between the two complex triplets is negligible, so $\chi_1^{\pm\pm}$ and $\chi_2^{\pm\pm}$ represent the two physical doubly charged scalars, $H_1^{\pm\pm} ,  H_2^{\pm\pm}$, in the theory. The results presented here focus on the mass hierarchy of the scalar sector in which only one singly charged scalar has a mass lower than the masses of the doubly charged scalars. Specifically, we treat $H_1^{\pm\pm}$ as the higher mass doubly charged scalar, with $H_2^{\pm\pm}$ having a lower mass than $H_1^{\pm\pm}$. Consequently, $H_2^{\pm\pm}$ decays into $\ell^\pm \ell^\pm$, $W^\pm W^\pm$, $W^\pm H_1^\pm$, and $H_1^\pm H_1^\pm$, where $H_1^\pm$ is the lightest singly charged scalar. With $< \chi_2^0 > = v_{\chi_2}$ and $\mathcal{O}^\pm$ denoting the matrix that rotates $(\phi^\pm, \chi_1^\pm, \xi^\pm, \chi_2^\pm)^T$ into $(G^\pm, H_1^\pm, H_2^\pm, H_3^\pm)^T$, the coupling strengths of $H_2^{\pm\pm}$ to $W^\mp W^\mp$ and $W^\mp H_1^\mp$ are given by,

\begin{eqnarray}
    g_{H_2^{\pm\pm} W^\mp W^\mp} &=& g v_{\chi_2} \\
    g_{H_2^{\pm\pm}(p_1) H_1^\mp(p_2) W^\mp} &=& \frac{2 m_w}{v} \mathcal{O}^{\pm}_{24}(p_1^\mu - p_2^\mu)
\end{eqnarray}

We denote the coupling strength of $H_2^{\pm\pm}$ to $H_1^\mp H_1^\mp$ by $\mu_2$ while the same coupling strength of $H_1^{\pm\pm}$ is denoted by $\mu_1$. 

The constraint coming from the $\rho$- parameter on such a scenario implies\cite{PDG},
\begin{equation}
3.0252~{\rm GeV}^2 \le v_\xi^2 - v_{T}^2 \le 8.7697~{\rm GeV}^2.
\end{equation}

with, $v_T$ being the effective complex triplet VEV given by,
\begin{equation}
v_{T}^2  = v_{\chi_1}^2 + v_{\chi_2}^2
\end{equation}

 In this scenario too the asymmetry generated under the resonant condition is given by, \eqref{ep}. We would like to emphasize that the goal of this article is not to propose a theoretical scenario. Instead, we have approached this framework purely from a phenomenological perspective, aiming to explore how a triplet VEV might be compatible with leptogenesis and what potential signatures could arise from this.  

\section{Results}
\label{res}

Due to the fact that we have considered resonant leptogenesis as the mechanism to generate the observed BAU, the leptogenesis scale is sufficiently low and here we focus specifically on the leptogenesis potential of a 500 GeV scalar. Around this energy scale, one should not neglect the mixing between different $SU(2)$ multiplets, which can potentially alter the results significantly. In order to point out the differences that this consideration brings, we will first demonstrate our results assuming $H_2^{\pm\pm}$ decays only to $\ell^{\pm}\ell^{\pm}$ and then show the results considering other decay modes.

This analysis is illustrative in character and is thus based on some simplifying assumptions, as mentioned below. 
\begin{enumerate}
    \item Mixing between two $SU(2)$ triplets, i.e, $\chi_1$ and $\chi_2$, is negligible while the effect of mixing between the doublet and triplet has been considered.
    
    \item The asymmetry generated by the heavier scalar, $H_1^{\pm\pm}$ is negligible and decays and inverse decays of the lighter scalar $H_2^{\pm}$  are entirely responsible for the generation of the observed BAU.
\end{enumerate}

Unlike right-handed neutrinos, the triplet scalars are not self-conjugates, and hence, there exists 4 Boltzmann equations instead of 2.

We denote with $n_p$ the number density of the type `$p$' particles.
Boltzmann equations describe the evolution of total number densities as a function of $z\equiv M_{H_2^{++}}/T = m_2/T$, with $m_2$ being the mass of the doubly charged scalar $H_2^{\pm\pm}$. The symmetric part of the number density of $H_2^{\pm\pm}$ is denoted by $\Sigma_{H_2}=(n_{H_2^{++}} + n_{H_2^{--}})/s$ 
and the asymmetric part is denoted by $\Delta_{H_2} = (n_{H_2^{++}} - n_{H_2^{--}})/s$. The asymmetry generated in the singly charged scalar is denoted by $\Delta_{H_1} = (n_{H_1^{+}} - n_{H_1^{-}})/s$ and the lepton asymmetry is denoted as $\Delta_L$. With these definitions, following Ref. \cite{hambye} the relevant Boltzmann equations are given by,

\begin{eqnarray}
 sHz\frac{d\Sigma_{H_2}}{dz}&=&-\left(\frac{\Sigma_{H_2}}{\Sigma^{eq}_{H_2}}-1\right)\gamma_D \nonumber \\
 \label{be1}
 sHz\frac{d\Delta_L}{dz}&=&\gamma_D\epsilon\left(\frac{\Sigma_{H_2}}{\Sigma^{eq}_{H_2}}-1\right)-2\gamma_DB_L \left(\frac{\Delta_L}{Y^{eq}_L}+\frac{\Delta_{H_2}}{\Sigma^{eq}_{H_2}}\right),
 \label{be2} \nonumber \\
 sHz\frac{d\Delta_{H_1}}{dz}&=&\gamma_D\epsilon\left(\frac{\Sigma_{H_2}}{\Sigma^{eq}_{H_2}}-1\right)-2\gamma_DB_
 {{H_1}}\left(\frac{\Delta_{{H_1}}}{Y^{eq}_{{H_1}}}-\frac{\Delta_{H_2}}{\Sigma^{eq}_{H_2}}\right),
 \label{be3} \nonumber \\
 sHz\frac{d\Delta_{H_2}}{dz}&=&-\gamma_D\left(\frac{\Delta_{H_2}}{\Sigma^{eq}_{H_2}}+B_L\frac{\Delta_L}{Y^{eq}_L}-B_{{H_1}}\frac{\Delta_{{H_1}}}{Y^{eq}_{{H_1}}}\right).
 \label{be4}
\end{eqnarray}

where $H$ is the Hubble constant at temperature $T$, 
$s$ is the total entropy density, $Y_X=n_X/s$,
a suffix $^{\rm eq}$ denotes equilibrium values, $\gamma_D$ is the space-time density of the total decay
process of ${H_2^{\pm\pm}}$ computed in thermal equilibrium.

\begin{equation}
    \gamma_D = s \Sigma_{H_2} \Gamma_D\frac{ k_1(z)}{k_2(z)}
\end{equation}
where $k_1$ and $k_2$ are the Boltzmann functions and $\Gamma_D$ is the total decay width of $H_2^{\pm\pm}$.

$B_L$ and $B_{H_1}$ are the branching ratios of ${H_2}^{\pm\pm}$ into $\ell^\pm \ell^\pm$ and $H_1^\pm H_1^\pm$ respectively.

Note that conservation of electric charge requires,

\begin{eqnarray}
    2\Delta_{H_2} + \Delta_{H_1} - \Delta_L = 0
\end{eqnarray}

The solution of Boltzmann equation \eqref{be2} is given by,
\begin{eqnarray}
    \Delta_L(z) &=& \Delta_i \exp({-\int_{z_{min}}^{z} \frac{2 \gamma_D B_L}{s Hz Y_L^{eq}}dz}) \notag \\
    &+& \int_{z_{min}}^z [ \gamma_D\epsilon\left(\frac{\Sigma_{H_2}}{\Sigma^{eq}_{H_2}}-1\right) - 2\gamma_DB_L \left(\frac{\Delta_{H_2}}{\Sigma^{eq}_{H_2}}\right)] \exp({-\int_{z^{\prime}}^{z} \frac{2 \gamma_D B_L}{s Hz Y_L^{eq}}dz})
\end{eqnarray}

Since there is no initial lepton asymmetry, $\Delta_i = 0$. As $\Sigma_{H_2}$ is the sole source to generate all the three asymmetries and $\Sigma_{H_2} > \Delta_{H_2}$, one may ignore the contribution of $\Delta_\chi$, the lepton asymmetry $\Delta_{H_2}$ scales with $\epsilon$.

Below, we show our results in terms of CP asymmetry $\epsilon$ needed to produce the observed BAU for a given set of $|\mu_2| , v_{\chi_2}$. Here we have taken the mass of ${H_2}^{++}$ to be 500 GeV. Since the sphaleron transition goes out of equilibrium around the energy scale of 200-100 GeV, the lepton asymmetry generated up to this temperature is converted to thebaryon asymmetry. This temperature corresponds to $z \approx 2.5-5$. In order to stay well above the electroweak phase transition, we have used $z=2.5$ for our further calculation.

\begin{figure}[htb!]
     \begin{subfigure}[h]{0.48\textwidth}
         \centering
         \includegraphics[width=1.1\textwidth,height=0.8\textwidth]{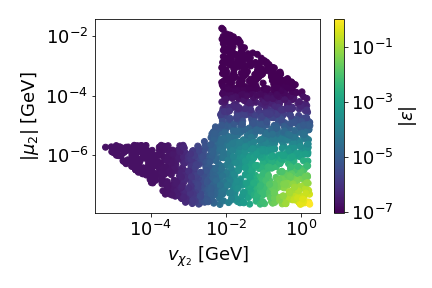}
         \caption{}
         \label{vtmu_app1_b4}
     \end{subfigure}
     \hfill
     \begin{subfigure}[h]{0.48\textwidth}
         \centering
         \includegraphics[width=1.1\textwidth,height=0.8\textwidth]{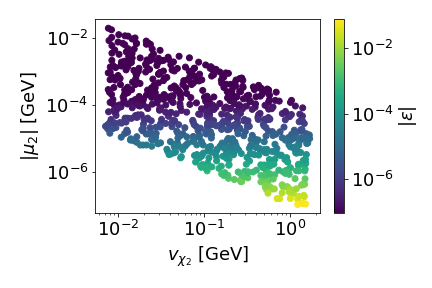}
         \caption{}
         \label{vtmu_app1_aftr}
     \end{subfigure}
     \begin{subfigure}[h]{0.48\textwidth}
         \centering
         \includegraphics[width=1.1\textwidth,height=0.8\textwidth]{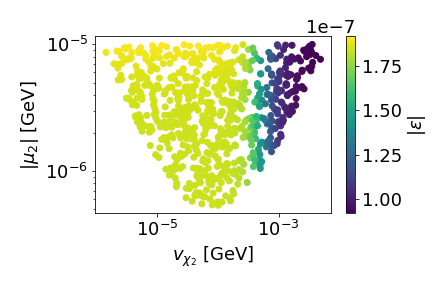}
         \caption{}
         \label{vtmu_app2_b4}
     \end{subfigure}
     \hfill
     \begin{subfigure}[h]{0.48\textwidth}
         \centering
         \includegraphics[width=1.1\textwidth,height=0.8\textwidth]{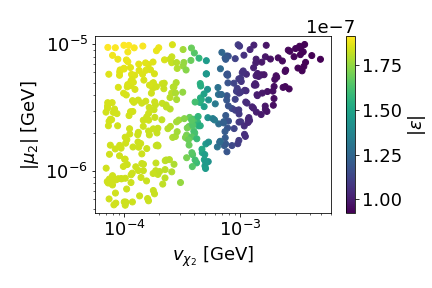}
         \caption{}
         \label{vtmu_app2_aftr}
     \end{subfigure}
    
        \caption{Top left: Values of $\epsilon$ needed to generate observed BAU for different values of $|\mu_2|$ and $v_{\chi_2}$, when only two decay modes of ${H_2}^{\pm\pm}$ i.e, ${H_2}^{\pm\pm} \to \ell^\pm \ell^\pm , H_1^\pm H_1^\pm$ has been taken into account. Top right: same as top left but with the constraint from the search for doubly charged scalar imposed. Bottom left: Values of $\epsilon$ needed to generate observed BAU for different values of $|\mu_2|$ and $v_{\chi_2}$, when all the decay modes of ${H_2}^{\pm\pm}$ i.e, ${H_2}^{\pm\pm} \to \ell^\pm \ell^\pm , W^\pm W^\pm , H_1^\pm W^\pm \ \text{and} \ H_1^\pm H_1^\pm$ has been taken into account. Bottom right: same as bottom left but with the constraint from the search for doubly charged scalar imposed. $\mathcal{O^\pm}_{24} = 10^{-6}$ in our considered parameter space. }
        \label{vtmu}
\end{figure}

The generated lepton asymmetry gets converted to baryon asymmetry via sphaleron transition. Following Ref. \cite{leptopedes} the baryon to photon ratio $({\eta_B})$ is related to lepton to photon ratio$(\eta_L)$ by,
\begin{equation}
    \eta_B = -\frac{a_{sph}}{f} \eta_L
\end{equation}
with $a_{sph} = 28/79$ \cite{chempot}, the conversion factor due to sphaleron transition and $f = 2387/86$, the dilution factor calculated assuming standard photon production from the onset of leptogenesis till recombination (taken from Ref. \cite{leptopedes}). With these numbers, 
\begin{equation}
    \eta_B = -0.013 \eta_L \approx 1.35 \Delta_L
\end{equation}

In the last step, we have used the relation, $s = g_{\star}n_{\gamma}$\cite{Kolb1990}, with $g_{\star} = 106$ and $n_\gamma$ is the number density of photon. $\eta_B = 6.1 \times 10^{-10}$ corresponds to $\Delta_L \approx 4.6 \times 10^{-10}$

As emphasized at the beginning of this section, We will demonstrate our results under two assumptions : (a). ${H_2}^{\pm\pm}$ decays only into $\ell^\pm \ell^\pm$ and $H_1^\pm H_1^\pm$. (b) ${H_2}^{\pm\pm}$ decays into $\ell^\pm \ell^\pm$ , $W^\pm W^\pm$, $W^\pm H_1^\pm$ and $H_1^\pm H_1^\pm$.

Since the mass of the doubly charged scalar, ${H_2}^{\pm\pm}$ is 500 GeV, there exists strong constraint imposed by the search of doubly charged scalar in the $\ell^{\pm} \ell^{\pm}$ decay mode which puts a lower limit of 450 GeV on the mass of the doubly charged scalar if the branching ratio of ${H_2}^{\pm\pm} \to \ell^{\pm}\ell^{\pm}$ exceeds 10\%.

For the given values of $\mu_2 \ \text{and} \ v_{\chi_2}$, we first obtain the maximum allowed value for $|\epsilon|$ which we denote as $|\epsilon|_{max}$ and check whether the thus generated lepton asymmetry, which we denote as $\Delta_{L_{max}}$, can serve to produce the observed BAU. If this lepton asymmetry predicts an under-abundance of net baryon number, the corresponding value of $\mu_2 \ \text{and} \ v_{\chi_2}$ is discarded. If the lepton asymmetry predicts an over-abundance of net baryon number, the values of $|\epsilon|$ were randomly scanned with an upper limit of $|\epsilon_{max}|$ to obtain the required value of $|\epsilon|$ that corresponds to observed BAU.

This value of $\epsilon$ then decides the permissible sets of $\mu_1, v_{\chi_1} \ \text{and} \ \Delta m^2$ which are consistent with the observed BAU. Here, we have taken the neutrino mass matrix to be diagonal with,

\begin{equation}
    m_\nu = m_\nu^1 + m_\nu^2 
\end{equation}

where $m_\nu^1 = Y_1 v_{\chi_1}$ and $m_\nu^2 = Y_2 v_{\chi_2}$, with $Y_1 = diag(y_{111} , y_{122}, y_{133}) \exp^{i\beta_2}$ and $Y_2 = y_2 diag(1,1,1) \exp^{i\beta_1}$. We have taken the VEV, $v_{\chi_1}$, and $v_{\chi_2}$, to be real while the Yukawa couplings remain complex. We consider $y_{211} = y_{222} = y_{233}$ and denote them as $y_2$, while $y_{111} \ne y_{122} \ne y_{133}$. Under these assumptions the two mass squared differences, i.e, ${\Delta m_{\nu_{12}}^2}$ and ${\Delta m_{\nu_{23}}^2}$ have been computed, and while obtaining the permissible sets of $\mu_1$ and $v_{\chi_1}$, only those values of $v_{\chi_1}, y_{111}, y_{122}, y_{133}$ have been used that satisfies the constraint on  ${\Delta m_{\nu_{12}}^2}$, and ${\Delta m_{\nu_{23}}^2}$ at $3\sigma$ level\cite{neu}. 
With $\beta_1 - \beta_2 = \beta$ and the relative phase between $\mu_1$ and $\mu_2$ denoted as $\alpha$, $\epsilon$ turns out to be,
\begin{equation}
\label{ep}
    \epsilon = \frac{[|\mu_1||\mu_2| \sin (\alpha - \beta) \Sigma_{k}|y_{1kk}||y_2|]}{8\pi^2(m_2^2-m_1^2)}\frac{m_2}{\Gamma_2}
\end{equation}

\begin{figure}[htb!]
     \begin{subfigure}[h]{0.48\textwidth}
         \centering
         \includegraphics[width=1.1\textwidth,height=0.8\textwidth]{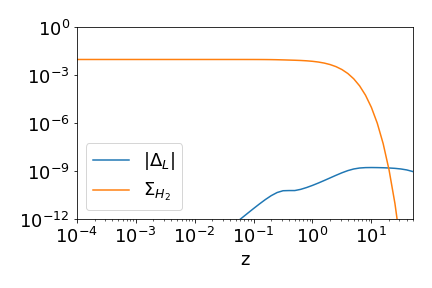}
         \caption{}
         \label{asym}
     \end{subfigure}
     \hfill
     \begin{subfigure}[h]{0.48\textwidth}
         \centering
         \includegraphics[width=1.1\textwidth,height=0.8\textwidth]{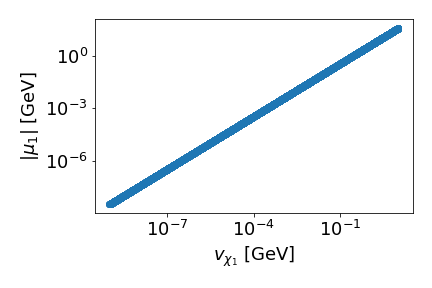}
         \caption{}
         \label{v1m1}
     \end{subfigure}
    
        \caption{Left: The abundances $\Sigma_{\chi_2}$ and $\Delta_L$ as a function of z for $m_2 = 500$ GeV , $|\mu_2| = 1.2 \times 10^{-6} $ GeV , $v_{\chi_2} = 3.9 \times 10^{-4} $ GeV, $|\epsilon| = 1.8 \times 10^{-7}$ and $\mathcal{O^\pm}_{24} = 10^{-6}$. Right: Corresponding values of $v_{\chi_1}$ and $\mu_1$. }
        \label{vt1mu1}
\end{figure}

For $m_2 = 500$ GeV , $|\mu_2| = 1.2 \times 10^{-6} $ GeV , $v_{\chi_2} = 3.9 \times 10^{-4} $ GeV and $|\epsilon| = 1.8 \times 10^{-7}$, the generated asymmetry considering all decay channels of $H_2^{\pm\pm}$ is shown in Fig. \ref{asym} and corresponding values of $\mu_1$ and $v_{\chi_1}$ is shown in Fig. \ref{v1m1}.

We point out below a few salient features extracted from Fig. \ref{vtmu}.

\begin{itemize}
    \item Fig. \ref{vtmu_app1_b4} shows the region of parameter space consistent with the observed BAU when ${H_2}^{\pm\pm}$ decays exclusively into $\ell^{\pm} \ell^{\pm}$ and $H_1^\pm H_1^\pm$. A lower bound on $v_{\chi_2}$ exists, as smaller values of $|v_{\chi_2}|$ result in a very small partial decay width of ${H_2}^{\pm\pm}$ into $H_1^\pm H_1^\pm$, which reduces $\epsilon_{max}$ to a level insufficient to account for the observed BAU.

    \item Fig. \ref{vtmu_app1_aftr} is similar to Fig. \ref{vtmu_app1_b4} but after imposing the constraint coming from the search for a doubly charged scalar in the $\ell^\pm \ell^\pm$ final state at the LHC. Since a lower value of $v_{\chi_2}$ corresponds to a high branching ratio of ${H_2}^{\pm\pm}$ into $\ell^\pm \ell^\pm$, which is disallowed by the above mentioned constraint, the allowed region shifts towards a relatively higher value of triplet VEV. This region of parameter space is also allowed by the search for a doubly charged scalar in $WW$ decay mode.

    \item Fig. \ref{vtmu_app2_b4} shows the region of parameter space consistent with the observed BAU when ${H_2}^{\pm\pm}$ decays into $\ell^{\pm} \ell^{\pm}$, $W^{\pm} W^{\pm}$,$W^{\pm} H_1^{\pm}$ and $H_1^\pm H_1^\pm$. In this case, $v_{\chi_2}$ within $10^{-6}~\text{GeV} - 10^{-2} \text{GeV}$ answers affirmatively to the observed BAU for ${\mu_2}$ within $5\times10^{-7}~\text{GeV} - 10^{-5} \text{GeV}$. The allowed value of $\mu_2$ is reduced with respect to Fig. \ref{vtmu_app1_b4} because, with all allowed decay channels available for $H_2^{\pm\pm}$, higher values of $\mu_2$ will suppress the branching ratio of $H_2^{\pm\pm}$ into $\ell^\pm \ell^\pm$ so much that sufficient lepton asymmetry will not be generated. 

     \item Fig. \ref{vtmu_app2_aftr} is the same as Fig. \ref{vtmu_app2_b4}, but here we have imposed the constraint coming from the search of doubly charged scalar in the $\ell^\pm \ell^\pm$ final state at the LHC. Again, this constraint rules out the region with low triplet VEV with a small portion of this region remaining corresponding to higher value of $\mu_2$. This happens because the higher value of $\mu_2$ serves to lower the branching ratio of ${H_2}^{\pm\pm}$ into $\ell^{\pm} \ell^{\pm}$ and hence the collider constraint is relaxed.  
\end{itemize}

\section{Summary and conclusion}
\label{cncl}

In this work, we have looked at the possibility of generating the observed BAU through leptogenesis in a way that is more generalized than earlier studies, allowing the leptogenesis scale to be within a few hundreds of GeV. A theoretical framework answering to leptogenesis, neutrino mass, and collider constraints has been discussed. The mass of the decaying scalar responsible for leptogenesis and the corresponding allowed triplet VEV satisfying all the constraints have been numerically illustrated. Although flavor constraints are not explicitly obtained, stronger constraints on the triplet VEV arise from the search for doubly charged scalar decaying into $WW$, thus making our study conservative.

\section{Acknowledgement}
The work of RG has been supported by a fellowship awarded by University Grants Commission India and the work of US is supported by a fellowship awarded by the Indian National Science Academy.

\end{document}